%% file: sn-article.tex
\documentclass[pdflatex,sn-jnl]{sn-jnl}

\usepackage{graphicx}
\usepackage{multirow}
\usepackage{amsmath,amssymb,amsfonts}
\usepackage{amsthm}
\usepackage{mathrsfs}
\usepackage[title]{appendix}
\usepackage[dvipsnames]{xcolor}
\usepackage{textcomp}
\usepackage{booktabs}
\usepackage{algorithm}
\usepackage{algpseudocode}
\usepackage{listings}
\usepackage{comment}
\usepackage{natbib}
\usepackage{subcaption}
\usepackage{placeins}
\usepackage{pdflscape}

\theoremstyle{plain}

\theoremstyle{remark}

\theoremstyle{definition}

\begin{document}

\title[Gamification Preferences in Education]{Gamification Preferences in Digital Education: The Role of Individual Differences}

\author[1]{\fnm{Anna Katharina} \sur{Ricker}}
\email{anna.ricker@kit.edu}

\author[1]{\fnm{Kai} \sur{Marquardt}}
\email{kai.marquardt@kit.edu}

\author*[1,2]{\fnm{Lucia} \sur{Happe}}
\email{lucia.happe@kit.edu}

\affil[1]{\orgdiv{Department of Informatics}, \orgname{Karlsruhe Institute of Technology}, \orgaddress{\street{Am Fasanengarten 5}, \city{Karlsruhe}, \postcode{76131}, \country{Germany}}}

\affil[2]{\orgdiv{Department of Information Management}, \orgname{University of Economics in Bratislava}, \orgaddress{\street{Dolnozemská cesta 1/b}, \city{Bratislava}, \postcode{85235}, \country{Slovakia}}}

\abstract{
Although personalization is widely advocated in gamified learning, empirical evidence on how learner characteristics and task context shape motivational preferences remains limited. This study examines how user characteristics and learning activity types relate to preferences for gamification elements in digital education. A large-scale quantitative survey (N = 530), including 34\% underage participants, assessed preferences for 13 gamification elements in relation to \textit{Age}, \textit{Gender}, HEXAD \textit{Player Type}, Big Five \textit{Personality Traits}, Felder–Silverman \textit{Learning Styles}, and Bloom-based \textit{Learning Activity Types}. Inferential statistical analyses and exploratory machine learning techniques revealed systematic but generally small-to-moderate effects across parameters. Age emerged as the most consistent predictor of preference, followed by player type and personality traits, whereas gender and learning styles showed comparatively weaker associations. In addition, learning activity type significantly influenced the perceived suitability of gamification elements, indicating that motivational design is task-dependent. The findings suggest that gamification effectiveness cannot be reduced to universally motivating elements. Instead, preferences are shaped by the interaction of learner characteristics and instructional context. These results provide empirical grounding for adaptive and modular gamification strategies in digital learning environments.
}



\maketitle

\input{Chapters/2.Introduction}

\input{Chapters/3.RelatedWork}
\input{Chapters/4.Methodology}

\input{Chapters/5.Results}
\input{Chapters/6.DiscussionConclusion}
\bibliographystyle{sn-basic}
\bibliography{sn-bibliography}

\begin{appendices}
\section{Questionnaires Used in This Study}
\label{sec:appendix:questionairs}
\FloatBarrier   

This section of the appendix contains the questionnaire items in all three language versions: English, German, and the simplified children’s version in German.
Basic demographic questions are omitted for brevity. 

For standardized instruments used in the questionnaire, such as \textit{Player Type}, \textit{Personality}, and \textit{Learning Style}, only the language versions newly created or adapted for this thesis are included here.

\begin{table}[!htb]
    \centering
    \begin{tabular}{|p{0.02\linewidth}|p{0.45\linewidth}|p{0.45\linewidth}|} \hline
    ~ & \textbf{German Version} & \textbf{Children Version} \\ \hline
    1 & Es macht mich glücklich anderen zu helfen. & Es macht mich froh anderen zu helfen. \\ \hline
    2 & Ich bin gerne Teil eines Teams. & Ich bin gerne Teil einer Gruppe \\ \hline
    3 & Es ist mir wichtig, meinen eigenen Weg zu gehen. & Es ist mir wichtig, meine eigenen Entscheidungen zu treffen. \\ \hline
    4 & Ich mag es, schwierige Aufgaben zu meistern. & Ich mag es, schwierige Aufgaben zu schaffen. \\ \hline
    5 & Ich würde mich als rebellisch bezeichnen. & Ich tue oft nicht das, was die Erwachsenen sagen. \\ \hline
    6 & Belohnungen sind ein tolles Mittel, um mich zu motivieren. & Belohnungen sind toll, um mich zu motivieren. \\ \hline
    7 & Das Wohlergehen anderer ist mir wichtig. & Mir ist wichtig, dass es anderen gut geht. \\ \hline
    8 & Gruppenaktivitäten machen mir Spaß. & Gruppenaufgaben machen mir Spaß. \\ \hline
    9 & Gelegenheiten zur Selbstentfaltung sind wichtig für mich. & Ich freue mich etwas neues zu entdecken, was mit Spaß macht. \\ \hline
    10 & Ich mag es, aus schwierigen Umständen siegreich hervorzugehen. & Ich mag es, schwierige Situationen zu gewinnen. \\ \hline
    11 & Ich halte mich nicht gerne an Regeln. & Ich halte mich nicht gerne an Regeln. \\ \hline
    12 & Bei angemessener Belohnung strenge ich mich gerne entsprechend an. & Wenn es eine passende Belohnung gibt, strenge ich mich gerne an. \\ \hline
    \end{tabular}
    \caption{Child and German Versions of the \textit{Player} Questionnaire}
    \label{fig:Questionaire:Player}
    \small Translation is based on the Short Version Questionaire by \cite{krathhexad-122023}.
\end{table}

\begin{table}[!htb]
    Ich... \par\medskip
    \centering
    \begin{tabular}{|p{0.45\linewidth}|p{0.45\linewidth}|} \hline
    \textbf{German Version} & \textbf{Children Version} \\ \hline
    1... bin eher zurückhaltend, reserviert. & ... bin eher ruhig und zurückhaltend. \\ \hline
    2... schenke anderen leicht Vertrauen, glaube an das Gute im Menschen. & ... vertraue anderen schnell und glaube, dass Menschen gut sind. \\ \hline
    3... bin bequem, neige zur Faulheit. & ... bin bequem und eher faul. \\ \hline
    4... bin entspannt, lasse mich durch Stress nicht aus der Ruhe bringen. & ... bin entspannt und bleibe auch bei Stress ruhig. \\ \hline
    5... habe nur wenig künstlerisches Interesse. & ... interessiere mich nicht so sehr für Kunst. \\ \hline
    6... gehe aus mir heraus, bin gesellig. & ... bin offen und gerne mit anderen zusammen. \\ \hline
    7... neige dazu, andere zu kritisieren. & ... sage anderen öfters, was sie besser machen können. \\ \hline
    8... erledige Aufgaben gründlich. & ... mache meine Aufgaben ordentlich. \\ \hline
    9... werde leicht nervös und unsicher. & ... werde schnell nervös und unsicher. \\ \hline
    10... habe eine aktive Vorstellungskraft, bin phantasievoll. & ... habe viel Fantasie und stelle mir gerne Dinge vor. \\\hline
    \end{tabular}
    \caption{Child Version of the \textit{Personality} Questionnaire}
    \label{fig:Questionaire:Personality}
    \small Translation is based on the short version by \cite{rammstedtmeasuring2007}; the German version is included for comparison.
\end{table}

\begin{landscape}
    \centering 
    \begin{table}[!htb]
    \centering
    \begin{tabular}{|p{0.3\linewidth}|p{0.3\linewidth}|p{0.3\linewidth}|} \hline
    \textbf{Englisch Version} & \textbf{German Version} & \textbf{Children Version} \\ \hline
    Tasks where you need to \textbf{remember} things, such as learning vocabulary. & Aufgaben bei denen du dich an Sachen \textbf{erinnern} sollst, wie beispielsweise beim Vokabeln lernen. & Aufgaben bei denen du dich an Sachen \textbf{erinnern} sollst, wie zum Beispiel beim Vokabeln lernen. \\ \hline
    Tasks where you need to \textbf{understand} things, such as summarizing a story to comprehend its content. & Aufgaben bei denen du Dinge \textbf{verstehen} sollst, wie beispielsweise indem ich eine Geschichte zusammenfasse, um den Inhalt zu verstehen. & Aufgaben bei denen du Dinge \textbf{verstehen} sollst, wie zum Beispiel indem du eine Geschichte zusammenfasst, um den Inhalt zu verstehen. \\ \hline
    Tasks where you need to \textbf{apply} what you have learned, such as using a math formula or answering quiz questions. & Aufgaben bei denen du Gelerntes \textbf{anwenden} sollst, wie beispielsweise die Anwendung einer Mathe Formel oder das Beantworten von Quiz-Fragen. & Aufgaben bei denen du das, was du schon gelernt hast \textbf{anwenden} sollst, wie zum Beispiel die Anwendung einer Mathe Formel oder das Beantworten von Quiz-Fragen. \\ \hline
    Tasks where you need to \textbf{analyze} concepts, such as classifying things based on specific characteristics or analyzing a poem. & Aufgaben bei denen du Konzepte \textbf{analysieren} sollst, wie beispielsweise das klassifizieren von Dingen basierend auf bestimmten Eigenschaften oder eine Gedichtsanalyse. & Aufgaben bei denen du über etwas \textbf{genau nachdenken} sollst, wie zum Beispiel das gruppieren von Tieren abhängig von dem Land aus dem sie kommen. \\ \hline
    Tasks where you need to \textbf{evaluate} something based on your knowledge, such as reviewing or assessing a classmate's/colleague's work. & Aufgaben bei denen du etwas \textbf{bewerten} sollst basierend auf deinem Wissen, wie beispielsweise das reviewen/bewerten einer Arbeit deines Mitschülers/Kollegen/Kommolitonen. & Aufgaben bei denen du etwas \textbf{bewerten} sollst, wie zum Beispiel das prüfen ob die Mathehausaufgaben deines Mitschülers richtig sind. \\ \hline
    Tasks where you need to \textbf{create} something yourself, such as writing a poem or a story. & Aufgaben bei denen du selbst etwas \textbf{erstellen} sollst, wie beispielsweise ein Gedicht oder eine Geschichte zu schreiben. & Aufgaben bei denen du selbst etwas \textbf{erstellen} sollst, wie zum Beispiel eine Geschichte zu schreiben. \\\hline
    \end{tabular}
    \caption{Questionnaire for \textit{Learning Activity Tasks} in All Languages}
    \label{fig:Questionaire:LAT}
    \small For each task, participants had to indicate whether the game element helped them, annoyed them, or neither.
    \end{table}
\end{landscape}

\begin{landscape}
    \centering 
    \begin{table}[!htb]
    \centering
    \begin{tabular}{|p{0.1\linewidth}|p{0.28\linewidth}|p{0.28\linewidth}|p{0.28\linewidth}|} \hline
    \textbf{\textit{Learning Style}} & \textbf{Item} & \multicolumn{2}{|c|}{\textbf{Alternatives}}  \\ \hline
    1 & In einem Buch mit vielen Bildern und Diagrammen schaue ich mir eher? & ... die Bilder und Diagramme genauer an. & ... den geschriebenen Text genauer an. \\ \hline
    2 & Ich erinnere mich am besten an? & ... das, was ich sehe. &  ... das, was ich höre. \\ \hline
    3 & Wichtiger ist mir, dass eine Lehrkraft...  &  ... den Stoff in klaren, aufeinander aufbauenden Schritten erklärt. &  ... einen Überblick gibt und Zusammenhänge zu anderen Themen herstellt. \\ \hline
    4 & Ich werde eher als jemand wahrgenommen, der? &  ... auf Details in der Arbeit achtet. &  ... kreativ an die Arbeit herangeht. \\ \hline
    5 & Ich würde lieber zuerst? &  ... Dinge ausprobieren. &  ... darüber nachdenken, wie ich es angehe. \\ \hline
    6 & Wenn ich eine Aufgabe erledigen muss, ziehe ich es vor? &  ... eine bewährte Methode zu meistern. &  ... neue Wege dafür zu finden. \\ \hline
    7 & Wenn mir Daten präsentiert werden, bevorzuge ich? &  ... Diagramme oder Grafiken. &  ... einen Text, der die Ergebnisse zusammenfasst. \\ \hline
    8 & Beim Schreiben einer Arbeit neige ich eher dazu? &  ... mit dem Anfang der Arbeit zu beginnen und dann chronologisch weiterzuschreiben. &  ... an verschiedenen Teilen zu arbeiten und sie später zu ordnen. \\ \hline
    9 & Wenn ich an einem Gruppenprojekt arbeiten muss, möchte ich zuerst? &  ... ein gemeinsames Brainstorming, bei dem alle Ideen einbringen. &  ... alleine brainstormen und danach die Ideen in der Gruppe vergleichen. \\ \hline
    10 & Wenn ich ein neues Thema lerne, bevorzuge ich...  & ... mich auf dieses Thema zu konzentrieren und so viel wie möglich darüber zu lernen. &  ... Verbindungen zwischen diesem Thema und verwandten Themen herzustellen. \\ \hline
    11 & Ich werde eher wahrgenommen als... &  ... kontaktfreudig. &  ... zurückhaltend. \\ \hline
    12 & Ich bevorzuge Kurse, die den Schwerpunkt auf? &  ... konkretes Material (Fakten, Daten). &  ... abstraktes Material (Konzepte, Theorien). \\ \hline
    \end{tabular}
    \caption{German Version of the \textit{Learning Style} Questionnaire}
    \label{fig:Questionaire:LearningStyleGERMAN}
    \small Translation is based on the Short Version Questionaire by \cite{godadevelopment2020}
    \end{table}
\end{landscape}

\begin{landscape}
    \centering 
    \begin{table}[!htb]
    \centering
    \begin{tabular}{|p{0.1\linewidth}|p{0.28\linewidth}|p{0.28\linewidth}|p{0.28\linewidth}|} \hline
    \textbf{\textit{Learning Style}} & \textbf{Item} & \multicolumn{2}{|c|}{\textbf{Alternatives}}  \\ \hline
    1 & In einem Buch mit vielen Bildern und Tabellen, schaue ich mir lieber... & ... die Bilder und Tabellen genauer an. & ... den Text genauer an. \\ \hline
    2 & Ich merke mir besser... & ... das, was ich sehe. & ... das, was ich höre. \\ \hline
    3 & Mir ist wichtig, dass eine Lehrerin oder ein Lehrer... & ... alles Schritt für Schritt erklärt. & ... einen Überblick gibt und zeigt, wie das Thema zu anderen Themen passt. \\ \hline
    4 & Andere denken, dass ich... & ... sehr genau arbeite. & ... gerne neue Ideen ausprobiere. \\ \hline
    5 & Ich würde lieber zuerst... & ... Dinge einfach ausprobieren. & ... überlegen, wie ich es machen will. \\ \hline
    6 & Wenn ich eine Aufgabe machen soll, mag ich lieber ... & ... es auf eine bekannte Weise zu machen. & ... neue Wege dafür zu finden. \\ \hline
    7 & Wenn mir jemand Zahlen oder Informationen zeigt, sehe ich sie lieber als... & ... Bilder, wie Tabellen oder Diagramme. & ... Text, der alles zusammenfasst. \\ \hline
    8 & Wenn ich eine Geschichte oder einen Text schreibe, schreibe ich lieber... & ... erst den Anfang und schreibe dann weiter. & ... verschiedene Teile und sortiere sie später. \\ \hline
    9 & Wenn ich mit anderen zusammen an einer Aufgabe arbeite, möchte ich zuerst ... & ... gemeinsam in der Gruppe Ideen sammeln. & ... alleine nachdenken und dann mit den anderen besprechen. \\ \hline
    10 & Wenn ich etwas Neues lerne, mag ich lieber... & ... nur dieses Thema lernen und viel darüber wissen. & ... schauen, wie es mit anderen Themen zusammenhängt. \\ \hline
    11 & Andere sehen mich eher als...  &  ... offen und gesprächig. &  ... ruhig und zurückhaltend. \\ \hline
    12 & Ich mag Unterricht, in dem es geht um... &  ... richtige Tatsachen und Daten. & ... Erklärungen, wie etwas funktioniert oder warum etwas passiert. \\ \hline
    \end{tabular}
    \caption{Child Version of the \textit{Learning Style} Questionnaire}
    \label{fig:Questionaire:LearningStyleCHILD}
    \small Translation is based on the Short Version Questionaire by \cite{godadevelopment2020}
    \end{table}
\end{landscape}
\FloatBarrier   

\section{Visualization and Description of the Taxonomy of Gamification Elements}
\label{sec:appendix:gamificationElements}
This section presents the visualization of all gamification elements used in the survey.
\begin{figure*}[htbp]
\centering

\begin{subfigure}{0.18\textwidth}
\centering
\includegraphics[width=\textwidth]{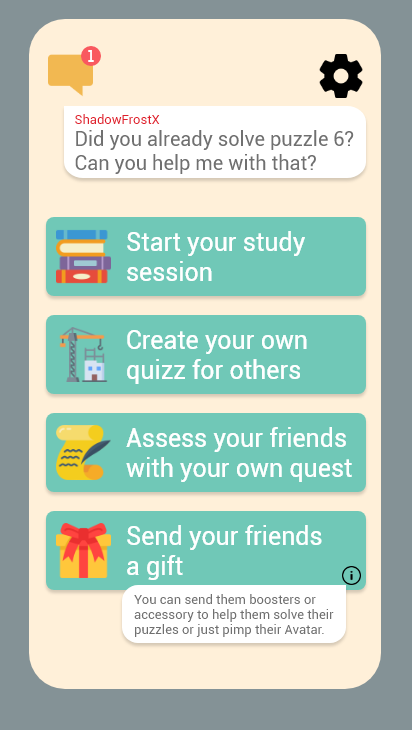}
\caption{\textit{Altruism}}
\end{subfigure}
\hfill
\begin{subfigure}{0.18\textwidth}
\centering
\includegraphics[width=\textwidth]{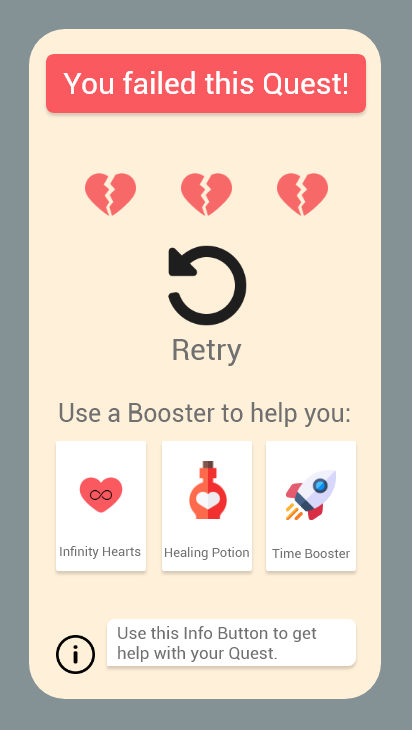}
\caption{\textit{Assistance}}
\end{subfigure}
\hfill
\begin{subfigure}{0.18\textwidth}
\centering
\includegraphics[width=\textwidth]{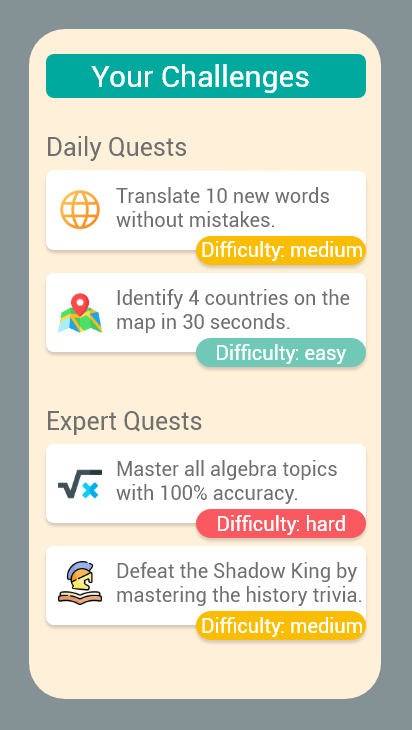}
\caption{\textit{Challenge}}
\end{subfigure}
\hfill
\begin{subfigure}{0.18\textwidth}
\centering
\includegraphics[width=\textwidth]{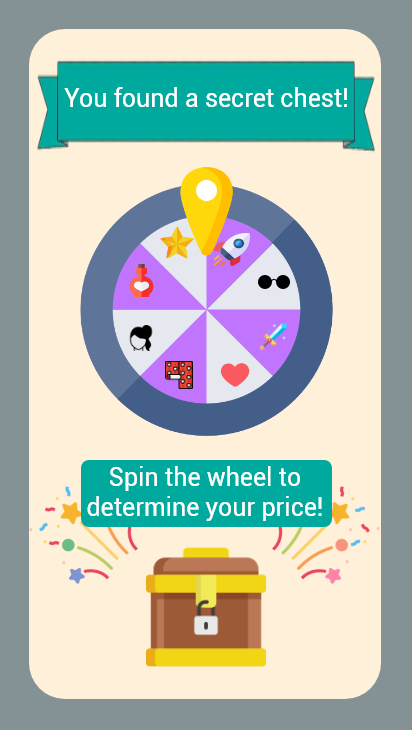}
\caption{\textit{Chance}}
\end{subfigure}
\hfill
\begin{subfigure}{0.18\textwidth}
\centering
\includegraphics[width=\textwidth]{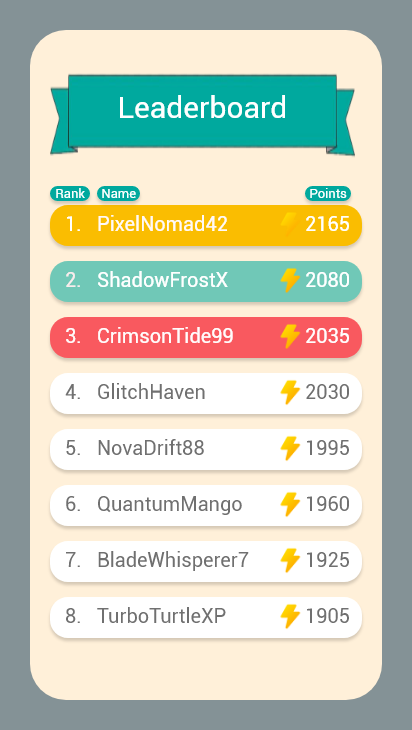}
\caption{\textit{Competition}}
\end{subfigure}

\vspace{0.2cm}

\begin{subfigure}{0.18\textwidth}
\centering
\includegraphics[width=\textwidth]{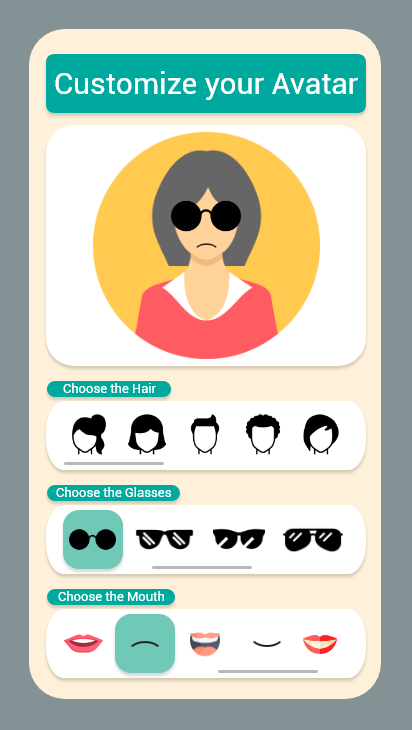}
\caption{\textit{Customis.}}
\end{subfigure}
\hfill
\begin{subfigure}{0.18\textwidth}
\centering
\includegraphics[width=\textwidth]{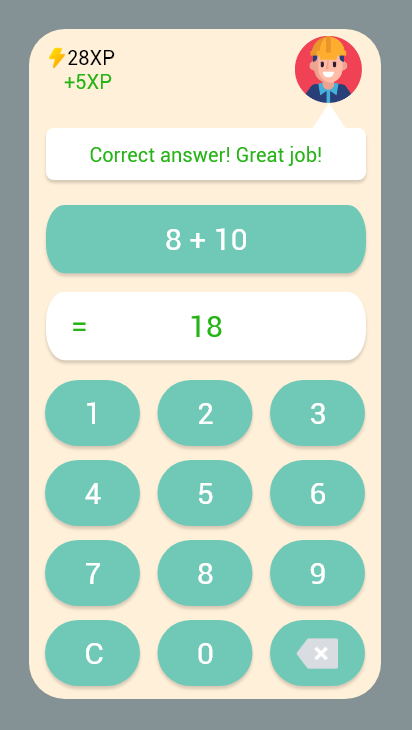}
\caption{\textit{Feedback}}
\end{subfigure}
\hfill
\begin{subfigure}{0.18\textwidth}
\centering
\includegraphics[width=\textwidth]{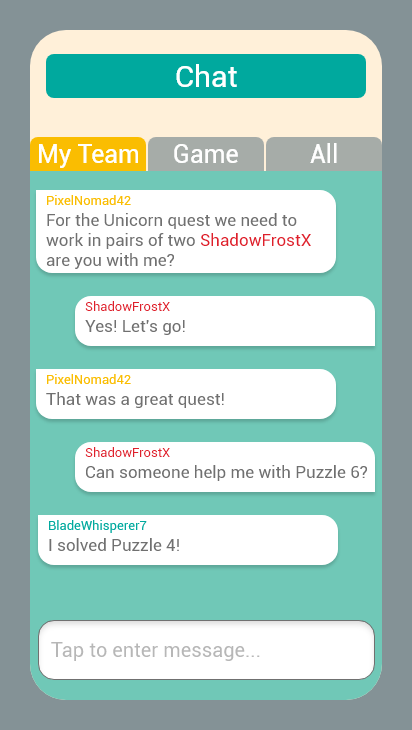}
\caption{\textit{Cooperation}}
\end{subfigure}
\hfill
\begin{subfigure}{0.18\textwidth}
\centering
\includegraphics[width=\textwidth]{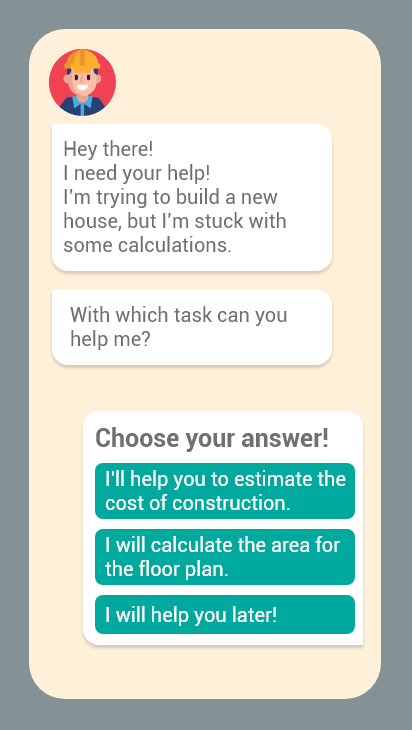}
\caption{\textit{Immersion}}
\end{subfigure}
\hfill
\begin{subfigure}{0.18\textwidth}
\centering
\includegraphics[width=\textwidth]{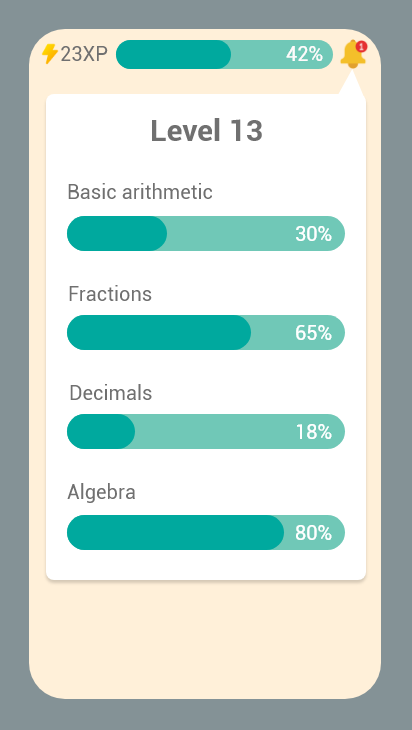}
\caption{\textit{Progression}}
\end{subfigure}

\vspace{0.2cm}

\begin{subfigure}{0.18\textwidth}
\centering
\includegraphics[width=\textwidth]{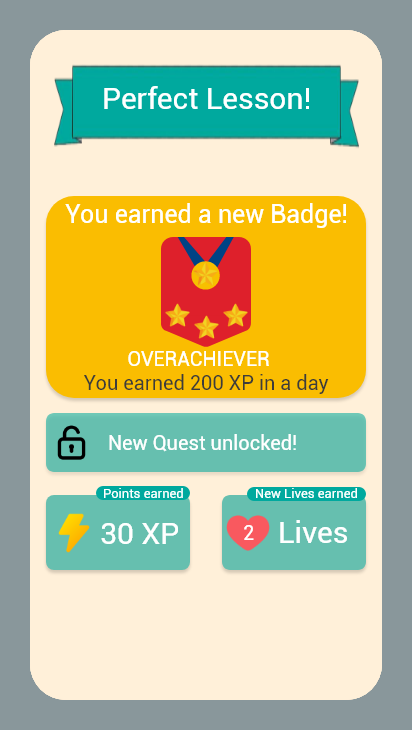}
\caption{\textit{Incentive}}
\end{subfigure}
\hfill
\begin{subfigure}{0.18\textwidth}
\centering
\includegraphics[width=\textwidth]{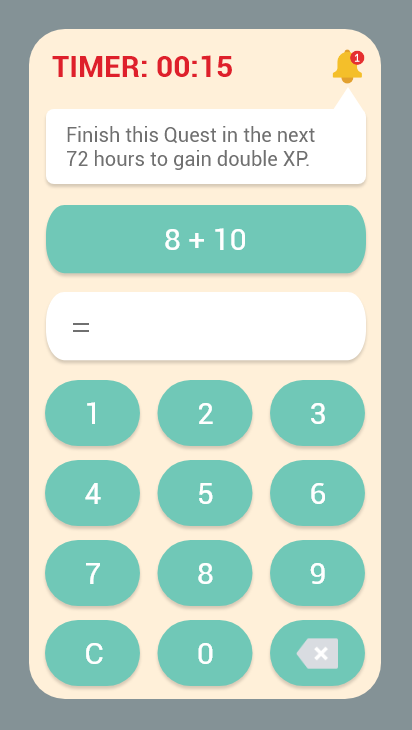}
\caption{\textit{TimePre.}}
\end{subfigure}
\hfill
\begin{subfigure}{0.18\textwidth}
\centering
\includegraphics[width=\textwidth]{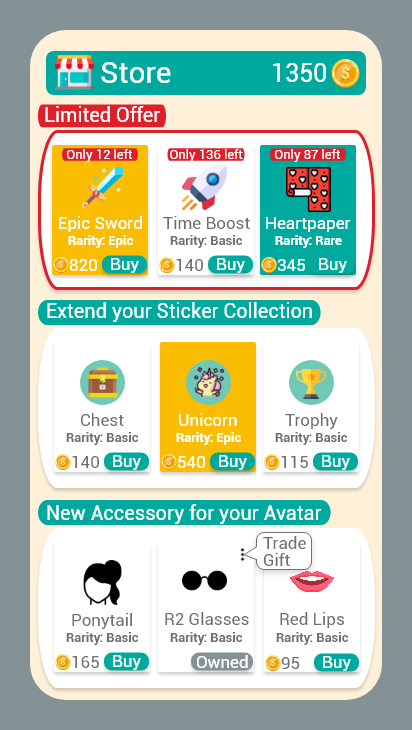}
\caption{\textit{Vir.Econ.}}
\end{subfigure}

\caption{Visualization of the 13 gamification elements used in the survey.}
\label{fig:gamificationTaxonomy}
\end{figure*}
\end{appendices}

\end{document}

%% file: Chapters/2.Introduction.tex
\section{Introduction}

Children are often criticized for lacking persistence in schoolwork, yet they frequently demonstrate remarkable dedication when engaging with games \cite{choupoints2015}. This contrast highlights the motivational potential of game mechanics and has inspired the integration of game design elements into non-game contexts through gamification. Gamification, the use of game design elements in non-game settings, has shown considerable promise for enhancing engagement, motivation, and learning outcomes in educational environments \cite{deterdinggame2011, cekerwhat2017, luiztailoring2023}. By embedding learning activities within structured game-like systems, educators aim to sustain attention and commitment, even during cognitively demanding tasks \cite{schobelcapturing2020}. Importantly, gamification supplements rather than replaces instructional content \cite{cekerwhat2017}.

Despite its potential, many educational platforms rely on a narrow set of commonly implemented elements such as Points, Badges, and Leaderboards. These elements are often treated as universally motivating due to their visibility and ease of implementation \cite{choupoints2015}. However, empirical findings increasingly suggest that gamification effects vary substantially depending on learner characteristics and contextual factors \cite{santosrelationship2021, luiztailoring2023}. In some cases, poorly aligned elements may even reduce engagement. This variability has led to growing interest in tailored gamification approaches that adapt design elements to individual traits or situational demands \cite{klocktailored2020}.

Existing research on tailored gamification, however, remains fragmented. Many studies focus on isolated relationships between a small number of elements and a single user characteristic, limiting comparability and generalizability. In addition, conceptual inconsistencies persist regarding how gamification elements are defined and grouped. Some investigations examine concrete implementations (e.g., badges), while others refer to broader motivational constructs (e.g., rewards), complicating synthesis across studies. Large-scale empirical investigations that integrate multiple user traits and learning context parameters within a coherent element taxonomy remain scarce.

Theoretically, individual differences in motivational responses may be explained by several complementary perspectives. Player typologies, such as the Hexad model, draw on self-determination theory to capture diverse motivational orientations toward autonomy, competence, and relatedness. Personality traits, particularly those captured by the Big Five framework, reflect stable dispositional tendencies that may influence how learners respond to challenge, competition, cooperation, or structure. Developmental factors such as age can shape motivational priorities and goal orientation, while pedagogical frameworks like Bloom’s Revised Taxonomy distinguish between cognitive task demands that may differentially align with specific game mechanics. Integrating these perspectives allows for a more comprehensive examination of how learner characteristics and instructional contexts jointly relate to perceived motivational suitability.

This study addresses these gaps through a large-scale quantitative user study (N = 530), including a substantial proportion of underage participants. Building on a structured taxonomy of gamification elements, the study simultaneously examines multiple learner characteristics, Age, Gender, Player Type, Personality Traits, and Learning Styles, alongside task contexts derived from Bloom’s Revised Taxonomy. Participants reported their individual traits and evaluated the motivational appeal of clustered gamification elements both in general and in relation to different types of learning activities.

The study is guided by the following research questions:

\begin{itemize}
\item[RQ1:] How do individual learner characteristics relate to perceived motivational appeal of different gamification elements?
\item[RQ2:] How do different types of learning activities influence the perceived suitability of gamification elements?
\item[RQ3:] Which learner and context factors emerge as comparatively stronger predictors of gamification preferences?
\end{itemize}

Rather than assuming universal effectiveness of specific elements, this work provides an empirically grounded mapping of how diverse learner profiles and instructional contexts relate to motivational preferences. By systematically comparing trait-based and task-based influences within a unified framework, the study contributes to a more nuanced understanding of individual differences in gamified education. These findings offer evidence-informed guidance for educators and instructional designers seeking to align motivational strategies with learner diversity and pedagogical goals.

%% file: Chapters/3.RelatedWork.tex
\section{Related Work}

Gamification research has increasingly moved from assuming universal effectiveness of design elements toward recognizing the importance of tailoring to individual and contextual differences. Early implementations often relied on widely recognizable mechanics such as points, badges, and leaderboards, implicitly assuming broad motivational appeal. However, accumulating evidence suggests that the effectiveness of gamification varies considerably across users and situations \cite{klocktailored2020, santosrelationship2021}. This has led to a growing body of work investigating how game elements can be adapted to specific user traits or contextual factors.

\subsection*{Trait-Based Tailoring}

A substantial portion of the literature focuses on tailoring gamification to user characteristics. Table~\ref{tab:relatedWorkuser} summarizes frequently investigated user-related parameters identified across 44 studies. Commonly examined factors include Player Type, Age, Gender, Personality Traits, Learning Style, and user motivation.

Among these, player typologies are particularly prominent. Originally developed in entertainment gaming contexts, models such as Bartle’s taxonomy \cite{bartleheartsnodate} and BrainHex \cite{nackebrainhexnodate} categorize players according to preferred play styles. However, these models were not explicitly designed for goal-oriented or educational systems. In contrast, Marczewski’s Hexad framework \cite{marczewskiuser2015} was developed specifically for gamified environments and is grounded in self-determination theory. It distinguishes six motivational orientations, Socializers, Free Spirits, Achievers, Philanthropists, Disruptors, and Players, reflecting different emphases on autonomy, competence, relatedness, and extrinsic rewards. For this reason, the Hexad model has become one of the most widely adopted frameworks in tailored gamification research \cite{klocktailored2020, tondelloelements2017} and serves as the primary player model in this study.

Beyond player typologies, personality traits have also been investigated as predictors of gamification preferences. The Big Five model (openness, conscientiousness, extraversion, agreeableness, neuroticism) conceptualizes personality along stable, dimensional traits and has been repeatedly linked to motivational tendencies and engagement patterns \cite{jiapersonality-targeted2016, hallifaxfactors2019}. Unlike categorical typologies, the Big Five framework captures gradual differences between individuals and may therefore offer complementary explanatory power in personalization contexts.

Demographic factors such as Age and Gender are frequently included in empirical studies, partly due to ease of measurement. However, findings remain inconsistent. Some studies report significant associations between demographic characteristics and preferred gamification elements \cite{klocktailored2020, luiztailoring2023}, whereas others find negligible or no differences \cite{kickmeier-rustgamification2014, thamrongratanalysis2020}. These mixed results highlight the need for larger, systematically structured investigations that consider demographic, psychological, and motivational traits simultaneously.

In educational contexts, Learning Style has also been proposed as a relevant tailoring parameter. The Felder–Silverman Learning Style Model (FSLSM) distinguishes learners along four dimensions (active–reflective, sensing–intuitive, visual–verbal, sequential–global) and is widely applied in technology-enhanced learning \cite{trajanovmodel2017}. While the predictive validity of learning styles in general remains debated, their continued use in adaptive educational systems justifies examining their relationship with gamification preferences.

\begin{table}[!ht]
    \centering
    \begin{tabular}{|p{0.3\linewidth} |p{0.6\linewidth}|}
    \hline
        \textbf{User Parameters} & \textbf{Included by} \\ \hline
        \raggedright Player Type &  \cite{klockdoes2018}, \cite{tondelloelements2017}, \cite{tondellodynamic2019}, \cite{tondellogamification2016}, \cite{hallifaxfactors2019}, \cite{palominoontology2022}, \cite{elgammaluser-based2020}, \cite{borgestowards2016}, \cite{conatiplayer2015}, \cite{oliveiratailored2019}, \cite{knutascreating2014}, \cite{santosrelationship2021}, \cite{moraquest2019}, \cite{bittencourttailor2020}, \cite{bovermanntowards2020}, \cite{oliveiradoes2020}, \cite{baldeonlega2016}
        \\ \hline   
        Age &  \cite{klockdoes2018}, \cite{tondelloelements2017}, \cite{tondellodynamic2019},  \cite{rodriguesautomating2022}, \cite{oyiboinvestigationnodate}, \cite{freundadvances2018},  \cite{marquardtgamification2024}, \cite{luiztailoring2023}, \cite{sprotenagenodate}, \cite{koivistodemographic2014}, \cite{oliveiratailored2019}, \cite{bittencourttailor2020}, \cite{thamrongratanalysis2020}, \cite{attaligamification2015}
        \\ \hline
         Gender &  \cite{klockdoes2018}, \cite{tondelloelements2017}, \cite{tondellodynamic2019},  \cite{rodriguesautomating2022}, \cite{oyiboinvestigationnodate}, \cite{freundadvances2018}, \cite{dendeninvestigation2017}, \cite{marquardtgamification2024}, \cite{luiztailoring2023}, \cite{sprotenagenodate}, \cite{koivistodemographic2014}, \cite{kickmeier-rustgamification2014}, \cite{oliveiratailored2019}\\ \hline
        \raggedright Personal traits, User types &  \cite{klockdoes2018}, \cite{tondelloelements2017}, \cite{tondellodynamic2019}, \cite{tondellogamification2016}, \cite{hallifaxfactors2019}, \cite{palominoontology2022}, \cite{jiapersonality-targeted2016}, \cite{elgammaluser-based2020}, \cite{codishacademic2014}, \cite{dendeneducational2017}, \cite{marache-franciscoprocess2013} \\ \hline   
        User motivation & \cite{attaligamification2015}, \cite{auerabout2021}, \cite{bittencourttailor2020}, \cite{hakulineneffect2014}, \cite{roostapersonalization2016}, \cite{klocktailored2020}, \cite{auvinenincreasing2015}, \cite{marache-franciscoprocess2013}
        \\ \hline
        Learning Style & \cite{baldeonlega2016}, \cite{aljabaliexperimental2005}, \cite{trajanovmodel2017}, \cite{zapalskalearning2006}, \cite{reiddigital2015} \\ \hline
        Usage frequency & \cite{klocktailored2020}, \cite{dendeninvestigation2017}, \cite{luiztailoring2023}, \cite{koivistodemographic2014}\\ \hline
        Culture, Ethnicity &  \cite{klocktailored2020}, \cite{marache-franciscoprocess2013}, \cite{todagamicsm2020}, \\ \hline
        Performance &  \cite{hallifaxfactors2019}, \cite{abramovichare2013}, \cite{baratarelating2014} \\ \hline
        \raggedright Preferred Game Genre/ Playing Setting & \cite{rodriguesautomating2022}, \cite{marquardtgamification2024}, \cite{luiztailoring2023} \\ \hline
        \raggedright Educational Level, Experience, Expertise & \cite{conatiplayer2015}, \cite{abramovichare2013}, \cite{reiddigital2015} \\ \hline
    \end{tabular}
    \caption{User Characteristics Considered in Tailored Gamification Literature}
   \label{tab:relatedWorkuser}
\end{table}

\subsection*{Context-Based Tailoring in Education}

Compared to user-based personalization, context-based tailoring has received less systematic attention. Context parameters examined in prior work include task type, academic subject, educational level, domain, and learning environment (Table~\ref{tab:relatedWorkcontext}). Among these, Learning Activity or Task Type is particularly relevant in educational settings, as different cognitive processes may demand distinct forms of motivational support.

Bloom’s Revised Taxonomy \cite{krathwohlrevision2002} offers a structured classification of cognitive processes ranging from remembering to creating. Several gamification studies have suggested that specific game elements may align differently with lower- versus higher-order cognitive tasks \cite{rodriguesautomating2022, luiztailoring2023}. For example, competitive elements may stimulate engagement in procedural tasks, whereas narrative or cooperative elements may better support complex analytical or creative activities. However, empirical evidence remains limited, and most investigations consider task context in isolation from learner traits.

\begin{table}[!ht]
    \centering
    \begin{tabular}{|p{0.3\linewidth} |p{0.6\linewidth}|}
    \hline
        \textbf{Context Parameters} & \textbf{Included by} \\ \hline
        \raggedright Learning Activity, Task & \cite{zapalskalearning2006}, \cite{marache-franciscoprocess2013}, \cite{palominoontology2022}, \cite{arnabmapping2015}, \cite{rodriguesautomating2022}, \cite{bovermanntowards2020}, \cite{baldeonlega2016}, \cite{dichevgamifying2017}, \cite{luiztailoring2023}, \cite{tondellorecommender2017} \\ \hline
        \raggedright Academic Subject & \cite{reiddigital2015}, \cite{dichevgamifying2017}, \cite{luiztailoring2023}, \cite{dichevagamification2014} \\ \hline
        \raggedright Educational Level & \cite{salmantailoring2024}, \cite{dichevgamifying2017}, \cite{luiztailoring2023}, \cite{dichevagamification2014} \\ \hline
        \raggedright Environment & \cite{rodriguesautomating2022}, \cite{auerabout2021}, \cite{marache-franciscoprocess2013} \\ \hline
        \raggedright Type of Application, Domain & \cite{dichevagamification2014}, \cite{hamarigamification2018}, \cite{dichevgamifying2017} \\ \hline
    \end{tabular}
    \caption{Context Parameters Considered in Tailored Gamification Literature}
   \label{tab:relatedWorkcontext}
\end{table}

\subsection*{Identified Gaps}

Although prior research has explored either user-based or context-based tailoring, few studies integrate multiple psychological models together with structured learning task classifications in a single large-scale investigation. Furthermore, inconsistencies in how gamification elements are defined, ranging from concrete mechanics (e.g., badges) to abstract motivational categories (e.g., rewards), limit cross-study comparability.

Large-scale empirical studies that combine psychometric profiling, demographic factors, and learning activity contexts within a coherent taxonomy of gamification elements remain scarce. By systematically integrating trait-based and task-based parameters within a unified framework, the present study seeks to provide a more comprehensive empirical foundation for evidence-informed gamification design in education.

%% file: Chapters/4.Methodology.tex
\section{Methodology}

\subsection{Design of Gamification Groups}

To conduct a large-scale preference study, a coherent and survey-feasible set of gamification elements was required. Rather than assuming that individual mechanics represent latent psychological constructs, we operationalized gamification elements as \emph{design clusters}—pragmatic groupings of commonly implemented mechanics intended to provide a standardized vocabulary for participant evaluation. This approach balances conceptual clarity with comprehensibility, particularly for younger participants.

An extensive literature review (70+ publications, including 17 systematic reviews) informed the identification and grouping of elements \cite{aljabaliexperimental2005, rodriguesautomating2022, tondelloelements2017, hallifaxfactors2019, oyiboinvestigationnodate, arnabmapping2015, jiapersonality-targeted2016, klocktailored2020, tondellogamification2016, trajanovmodel2017, moraquest2019, santosrelationship2021, bittencourttailor2020, freundusing2018, marquardtgamification2024, oliveiratailored2019, dendeninvestigation2017}. 
Given that prior work describes more than 50 distinct elements, clustering was necessary to keep the instrument manageable.

Building on the exploratory factor analysis of \citet{tondelloelements2017}, the classification was revised and expanded, resulting in 13 element groups:

\begin{itemize}
    \item \textbf{Altruism} allows one to make a meaningful contribution to other users or the app itself by sharing knowledge or items with others or helping to improve the system.
    \item \textbf{Assistance} supports you in your tasks through, for example, guides, power-ups and boosters, the ability to retry a task, or free mistakes.
    \item \textbf{Challenge} strategically challenges you through, for example, riddles, puzzles, difficult tasks, or so-called boss fights that test all the knowledge you have gained so far.
    \item \textbf{Chance} enables completely random outcomes, primarily in the form of rewards through, for example, spinning wheels or mystery boxes.
    \item \textbf{Competitions} are, for example, leaderboards, as well as duels or contests where you compete against your fellow players.
    \item \textbf{Cooperation} or Guild allows you to complete tasks as a team with your fellow players or communicate with them via chat.
    \item \textbf{Customization} allows you to design your own digital avatar or customize the game world according to your personal preferences.
    \item \textbf{Feedback} provides you with positive or negative visual or auditory responses regarding your behavior.
    \item \textbf{Immersion} gives tasks meaning through storytelling, narrative, exploration, and choices. 
    \item \textbf{Incentives} or Rewards are elements like badges, achievements, collectibles, or points that you receive, for example, for completing tasks.
    \item \textbf{Progression} displays your current overall or specific progress through for example levels, milestones, progress bars, or concept maps.
    \item \textbf{Time Pressure} aims to motivate one through time pressure, with elements such as a time limit or a given deadline.
    \item \textbf{Virtual Economy} allows you to buy, trade, or gift virtual items using virtually earned currency.
\end{itemize}
Several modifications were introduced. Time-related mechanics were separated into a distinct \textit{Time Pressure} cluster, as timers and deadlines are common in educational gamification but underrepresented in prior classifications. The former “Risk/Reward” factor was divided into \textit{Challenge} (mastery-oriented mechanics) and \textit{Chance} (luck-based reward mechanics) to distinguish skill-based from stochastic motivational drivers. \textit{Progression} and \textit{Feedback} were separated to reflect the conceptual difference between long-term goal advancement and immediate response to user action \cite{klocktailored2020}. The former \textit{Socialization} cluster was refined into \textit{Competition}, \textit{Cooperation}, and \textit{Virtual Economy} to distinguish competitive interaction, collaborative interaction, and exchange-based systems.

The resulting taxonomy aims to balance comprehensiveness, conceptual coherence, and survey feasibility.

\subsection{Study Design}

The survey was conducted online between March and May 2025 using LimeSurvey. Participation was open to individuals aged 8 and above who were proficient in German or English. A total of \textbf{N = 530} valid responses were included in the final dataset.

Three survey versions were provided: German, English, and a simplified German version designed for primary school children. The simplified version preserved the meaning of validated instruments while reducing linguistic complexity. All versions underwent pilot testing using the Think-Aloud method with an expert, a layperson, an educator, and a nine-year-old child to ensure clarity and feasibility.

The questionnaire comprised the following sections:

\begin{enumerate}
    \item Demographics (Age, Gender, Nationality)
    \item Player Type (12-item short HEXAD version \cite{krathhexad-122023})
    \item Personality (10-item Big Five short version \cite{rammstedtmeasuring2007})
    \item General preference ratings for 13 gamification element groups (5-point Likert scale from “demotivates me a lot” to “motivates me a lot”)
    \item Favorite element (single choice + open text explanation)
    \item Task-based element suitability (6 Bloom cognitive levels; 3-point scale: helpful / neutral / annoying)
    \item Learning Style (optional; short FSLSM version \cite{godadevelopment2020})
\end{enumerate}

Validated short scales were used for HEXAD and Big Five to maintain survey compactness. Where available, validated German translations were employed. The FSLSM scale was translated manually due to the absence of a validated German short version; careful forward translation and pilot testing were conducted to ensure comprehensibility.

Internal consistency (Cronbach’s $\alpha$) was calculated for multi-item scales for the full sample and separately for minors and adults.

For task-based preferences, three element groups (Customization, Virtual Economy, Chance) were excluded, as they were considered less directly task-dependent and more context-independent, making meaningful mapping to specific cognitive processes difficult in a brief survey format.

\subsection{Recruiting}
Participation was voluntary and anonymous. Before beginning the survey, participants received an explanation of the study's purpose and provided informed consent. For minors, parental consent and child assent were obtained prior to participation.
Adult participants were recruited primarily via social media, email distribution lists, and messaging platforms. Recruitment of children was conducted in cooperation with primary and secondary schools. In primary schools, survey sessions were conducted during regular class time with researcher assistance. In secondary schools, participation was supervised by teachers.
The study was conducted in accordance with applicable data protection regulations.



\subsection{Data Preperation and Analysis Plan}

After excluding incomplete entries and implausible responses (e.g., uniform answers across all personality or player type items, reported ages below 6 or above 80), the final dataset included \textbf{N = 530} participants. Of these, 344 were adults and 186 were minors (primary and secondary school students). The median completion time was 17.5 minutes (adults: 14 minutes; children: 18.5 minutes).

\subsubsection{Data Preparation}

Survey versions for adults and children were merged into a unified dataset. Column names and response formats were standardized prior to analysis. 

Scale scores were computed according to established scoring procedures. For HEXAD and Big Five, mean scale values were calculated for each participant. Learning style dimensions were derived from the respective short-scale items; tied or incomplete dimensions were coded as missing. 

Free-text responses were stored separately and analyzed qualitatively but are not the focus of the present quantitative analysis.

\subsubsection{Statistical Analysis}

The dependent variables were preference ratings for the 13 gamification element groups and task-based suitability ratings. Independent variables included \textit{Age}, \textit{Gender}, \textit{Player Type}, \textit{Personality Traits}, \textit{Learning Style}, and \textit{Learning Activity Task}.

Analyses were conducted in Python in two stages. First, descriptive statistics (means, standard deviations, distributions) were computed to summarize central tendencies and variability. Second, inferential analyses examined associations between user/context parameters and gamification preferences. Descriptive comparisons between children and adults were conducted; however, inferential results are reported for the combined sample unless stated otherwise. Likert-scale ratings were treated as interval-scaled for inferential analysis, consistent with common practice in large-sample survey research.

Independent samples t-tests were used for binary variables (e.g., gender). Pearson correlations were applied for continuous predictors (age, personality traits, player types), with Spearman correlations calculated where assumptions were partially violated. One-way ANOVAs with Tukey post-hoc comparisons were conducted for multi-category variables (e.g., learning activity tasks). Effect sizes were reported using Cohen’s $d$ and eta squared ($\eta^2$). Effect sizes were interpreted according to conventional thresholds (Cohen’s d: 0.2 small, 0.5 medium, 0.8 large; $\eta^2$: 0.01 small, 0.06 medium, 0.14 large).

To reduce Type I error due to multiple comparisons, p-values were adjusted using the Benjamini–Hochberg procedure. To control for multiple testing across element–parameter comparisons, p-values were adjusted within test families using the Benjamini–Hochberg false discovery rate (FDR) procedure. FDR was preferred over Bonferroni to balance Type I and Type II error in large comparison families.

To explore potential non-linear age effects, k-means clustering was applied to derive age groups, followed by ANOVA analyses. In addition, a Random Forest model was used to estimate the relative importance of predictor variables. These analyses are interpreted as exploratory. 

Assumptions of normality and homogeneity of variance were evaluated using Shapiro–Wilk and Levene’s tests. Where violations occurred, Mann–Whitney U tests were additionally computed to ensure robustness. Given the sufficiently large group sizes (n > 30), parametric tests were retained as primary analyses in accordance with the central limit theorem.

%% file: Chapters/5.Results.tex
\section{Results}

\subsection{Overall Preferences}

\begin{figure} [!htb]
  \centering
  \includegraphics[width=\linewidth]{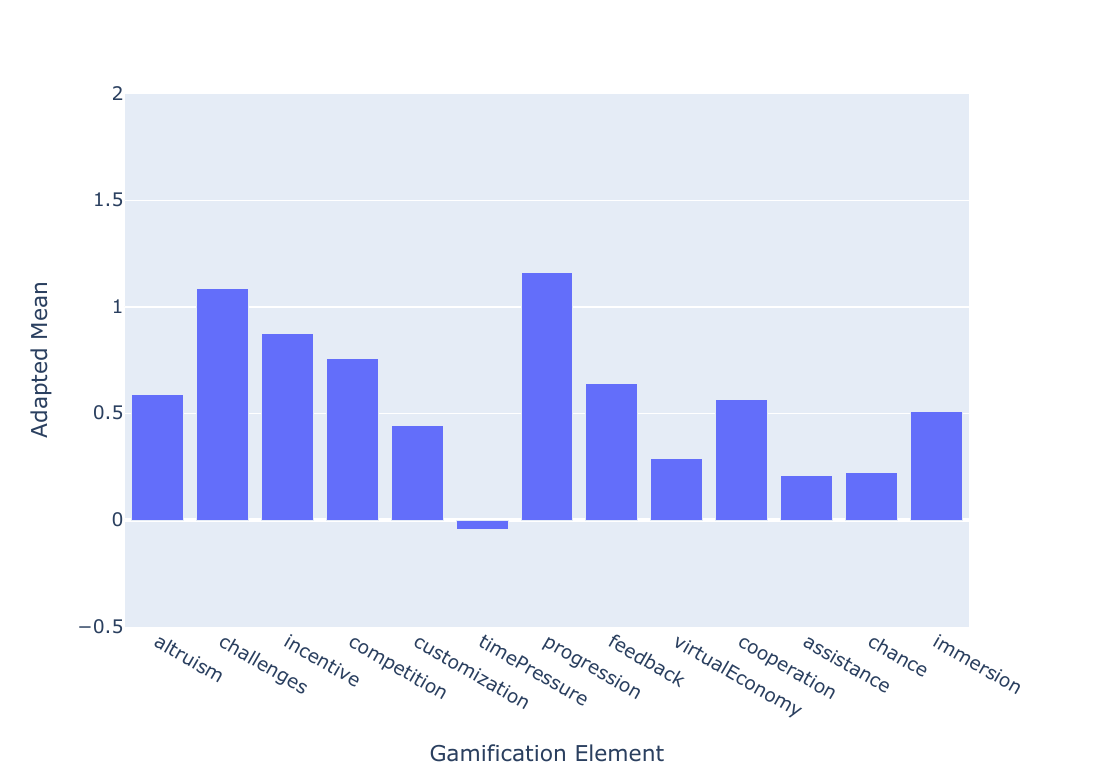}
  \caption{Mean preference ratings for each gamification element group (overall sample).}
  \label{fig:descriptiveGamificationElementsAll}
\end{figure}

To contextualize parameter-specific effects, we first report overall preference patterns across the full sample (5-point Likert scale from \emph{demotivates me a lot} to \emph{motivates me a lot}). Figure~\ref{fig:descriptiveGamificationElementsAll} summarizes mean ratings across the 13 element groups. Preferences clustered strongly around a small set of elements: \textit{Progression} received the highest mean rating, followed by \textit{Challenge}, \textit{Incentives}, and \textit{Competition}. In contrast, \textit{Time Pressure} was rated lowest across the sample.

When asked to select a single favorite element, \textit{Incentives} and \textit{Competition} exchanged rank compared to the mean-rating overview, suggesting that \textit{Competition} is more polarizing (high for some participants, but not broadly preferred), whereas \textit{Incentives} are more consistently selected as a top choice.

\subsection{Gender Preferences}

Gender differences in gamification element preferences were examined using two-sided independent-samples t-tests. Participants who did not self-identify as female or male (n = 8) were excluded from inferential comparisons due to small subgroup size but retained in descriptive summaries.

Across the 13 elements, observed gender differences were generally small in magnitude (all Cohen’s $d < .40$). After controlling for multiple comparisons using the Benjamini–Hochberg false discovery rate (FDR) procedure within this test family, statistically significant effects remained for \textit{Assistance}, \textit{Feedback}, and \textit{Chance}. In all three cases, women reported higher preference ratings than men. Effect sizes were small to approaching small-to-moderate (largest for \textit{Assistance}, $d = .36$).

\textit{Competition} showed a small effect in the opposite direction (higher ratings among men; $d = .18$), but this effect did not remain robust after FDR correction and is therefore interpreted cautiously.

To account for mild non-normality in several element distributions (skew around the neutral midpoint), Mann–Whitney U tests were additionally computed. The pattern of results was consistent for the elements that remained significant after correction, supporting the robustness of the findings.

Overall, gender-related differences were present but modest in size, indicating that while certain supportive elements (e.g., \textit{Assistance}, \textit{Feedback}) were more strongly endorsed by women, gender alone explains only a limited proportion of variance in gamification preferences.

\begin{table}[!htb]
\centering
\small
\begin{tabular}{l c c c c c}
\hline
\textbf{Element} & \textbf{$t$} & \textbf{$p$} & \textbf{$p_{FDR}$} & \textbf{$d$} & \textbf{$p_{U}$} \\
\hline
Altruism        & 0.04  & .967 & .967 & 0.00 & .902 \\
Challenge       & -0.65 & .515 & .670 & 0.06 & .349 \\
Incentive       & -0.84 & .402 & .603 & 0.07 & .366 \\
Competition     & 2.01  & .045 & .089 & 0.18 & .033 \\
Customization   & -2.73 & .006 & .017 & 0.24 & .005 \\
Time Pressure   & -0.34 & .730 & .781 & 0.03 & .715 \\
Progression     & -0.81 & .416 & .603 & 0.07 & .249 \\
Feedback        & -3.14 & .002 & .007* & 0.28 & .001 \\
Virtual Economy & 1.13  & .258 & .419 & 0.10 & .166 \\
Cooperation     & -0.73 & .467 & .603 & 0.06 & .471 \\
Assistance      & -4.05 & $< .001$ & $< .001$* & 0.36 & $< .001$ \\
Chance          & -3.27 & .001 & .005* & 0.29 & .002 \\
Immersion       & -0.41 & .683 & .781 & 0.04 & .730 \\
\hline
\end{tabular}
\caption{Independent-samples t-tests for gamification preferences by gender (two-sided). 
$p_{FDR}$ values reflect Benjamini–Hochberg correction within this test family. 
Positive $t$ indicates higher ratings among men; negative $t$ indicates higher ratings among women.}
\label{fig:genderTTest}
\end{table}




\subsection{Learning Style Preferences}


Learning style differences were examined across the four FSLSM dimensions: \textit{Intuitivity} (Intuitive vs. Sensor), \textit{Perception} (Visual vs. Verbal), \textit{Processing} (Active vs. Reflective), and \textit{Understanding} (Global vs. Sequential). Independent-samples t-tests were conducted for each element within each dimension (Table \ref{tab:ls_all_td}). Due to subgroup size imbalances (e.g., Visual $n=300$ vs. Verbal $n=47$), results should be interpreted cautiously. False discovery rate (FDR) correction was applied within each learning-style dimension.

Across dimensions, effect sizes were predominantly small ($d < .30$), with a few small-to-moderate effects in the \textit{Processing} dimension.

For \textit{Intuitivity}, Intuitive learners reported higher preferences for \textit{Chance}, \textit{Virtual Economy}, and \textit{Altruism}, whereas Sensor learners favored \textit{Progression}. Effect sizes were small, with \textit{Chance} showing the strongest association. Most effects remained stable in non-parametric robustness checks, although the difference for \textit{Progression} was less consistent.

For \textit{Perception}, only \textit{Progression} showed a significant difference, with Visual learners rating it higher than Verbal learners. No other reliable differences emerged, and given the strong subgroup imbalance, this dimension showed limited explanatory power.

The clearest pattern appeared in the \textit{Processing} dimension. Active learners showed higher preferences for \textit{Competition}, \textit{Cooperation}, \textit{Chance}, and \textit{Virtual Economy}. Effects for \textit{Competition} and \textit{Cooperation} were small to moderate ($d \approx .40$), representing the largest learning-style-related differences observed in the study.

For \textit{Understanding}, Global learners rated \textit{Challenge} slightly higher than Sequential learners; however, the effect size was small and subgroup imbalance warrants cautious interpretation.

Overall, the \textit{Processing} dimension (Active vs. Reflective) demonstrated the most consistent and comparatively stronger associations with gamification preferences. Other learning-style dimensions showed weaker and less stable patterns.

\begin{table}[ht]
\centering
\small
\begin{tabular}{l cc cc cc cc}
\hline
\textbf{Element} &
\multicolumn{2}{c}{\textbf{Intuitivity}} &
\multicolumn{2}{c}{\textbf{Perception}} &
\multicolumn{2}{c}{\textbf{Processing}} &
\multicolumn{2}{c}{\textbf{Understanding}} \\
 & $t$ & $d$ & $t$ & $d$ & $t$ & $d$ & $t$ & $d$ \\
\hline
Altruism        &  2.26*  & 0.25 & -0.84 & 0.13 &  1.70 & 0.19 & -0.59 & 0.07 \\
Challenge       &  0.58   & 0.07 &  0.12 & 0.02 &  0.16 & 0.02 &  2.16* & 0.25 \\
Incentive       &  0.82   & 0.09 &  0.68 & 0.11 & -0.38 & 0.04 & -1.69 & 0.19 \\
Competition     &  1.33   & 0.15 & -0.10 & 0.02 &  3.83** & 0.42 & -1.12 & 0.12 \\
Customization   &  1.53   & 0.17 & -1.47 & 0.23 &  1.77 & 0.19 & -1.65 & 0.19 \\
Time Pressure   & -0.51   & 0.06 &  0.74 & 0.12 &  0.54 & 0.06 & -0.02 & 0.00 \\
Progression     & -2.05*  & 0.23 &  2.45* & 0.36 &  0.16 & 0.02 &  1.52 & 0.17 \\
Feedback        & -0.98   & 0.11 &  0.25 & 0.04 & -0.75 & 0.08 &  1.19 & 0.13 \\
Virtual Economy &  2.52*  & 0.28 & -0.75 & 0.12 &  2.64** & 0.29 & -0.98 & 0.11 \\
Cooperation     &  0.75   & 0.09 &  1.23 & 0.19 &  3.53** & 0.39 &  0.61 & 0.07 \\
Assistance      & -0.26   & 0.03 & -0.02 & 0.00 &  1.15 & 0.13 & -0.30 & 0.03 \\
Chance          &  2.88** & 0.33 & -0.58 & 0.09 &  2.66** & 0.29 & -1.04 & 0.12 \\
Immersion       &  1.15   & 0.13 & -1.72 & 0.27 & -0.43 & 0.05 & -0.20 & 0.02 \\
\hline
\end{tabular}
\caption{Independent-samples t-tests (two-sided) and effect sizes (Cohen’s $d$) for gamification preferences across FSLSM dimensions. 
Positive $t$ indicates higher ratings for Intuitive, Visual, Active, and Global learners respectively. 
* $p < .05$, ** $p < .01$ (FDR-corrected within each learning-style dimension).}
\label{tab:ls_all_td}
\end{table}

\subsection{Player Type Preferences}

The HEXAD framework was analyzed due to its direct relevance to gamification design. In the present sample, \textit{Philanthropists} (n = 264) were most prevalent, followed by \textit{Accomplishers} (n = 183), \textit{Free Spirits} (n = 178), \textit{Players} (n = 131), and \textit{Socializers} (n = 114), while \textit{Disruptors} were underrepresented (n = 19), consistent with prior findings \cite{tondelloelements2017}.  

Associations between player types and gamification preferences were examined using Pearson correlations, with Spearman rank correlations computed as robustness checks. Given the number of comparisons, p-values were adjusted using the Benjamini–Hochberg false discovery rate procedure.

Overall, correlations were predominantly small in magnitude. No stable associations emerged for the \textit{Disruptor} type, and the element \textit{Immersion} did not show consistent relationships with any player type.

\textit{Philanthropists} showed consistent positive associations with \textit{Altruism}, \textit{Feedback}, \textit{Cooperation}, and \textit{Assistance}, aligning with their prosocial orientation.  
\textit{Socializers} correlated most strongly with \textit{Cooperation}, representing the largest effect observed in this analysis.  
\textit{Accomplishers} demonstrated a clear association with \textit{Challenge}, alongside smaller relationships with \textit{Competition} and \textit{Progression}.  
\textit{Players} showed the strongest association with \textit{Incentive}, and additional links to \textit{Competition}, \textit{Virtual Economy}, and \textit{Chance}.  
\textit{Free Spirits} displayed weaker but statistically supported preferences for elements such as \textit{Altruism} and \textit{Virtual Economy}.

While most effects were small, three comparatively stronger and theoretically coherent associations stood out: \textit{Socializer–Cooperation}, \textit{Accomplisher–Challenge}, and \textit{Player–Incentive}. These findings align with HEXAD theory and support its suitability as a basis for tailoring gamification design.

\begin{table}[ht]
\centering
\small
\begin{tabular}{l c c c c c c}
\hline
\textbf{Element} & \textbf{Philanth.} & \textbf{Social.} & \textbf{Free Sp.} & \textbf{Accompl.} & \textbf{Disrupt.} & \textbf{Player} \\
\hline
Altruism        & .21** & .13* & .21** & .02  & .03  & .06  \\
Challenge       & .13*  & .10* & .09*  & .30** & -.06 & .02  \\
Incentive       & .11*  & .07  & .10*  & .05  & -.06 & .31** \\
Competition     & .04   & .14**& .03   & .20** & .04  & .15** \\
Customization   & .05   & .03  & .09*  & -.05 & .02  & .09*  \\
Time Pressure   & .01   & .10* & .06   & .10* & -.02 & .03   \\
Progression     & .10*  & .06  & .04   & .20** & -.08 & .09*  \\
Feedback        & .17** & .11* & .03   & .02  & -.00 & .02   \\
Virtual Economy & .02   & .05  & .13** & .03  & .05  & .14** \\
Cooperation     & .16** & .29**& .09*  & .01  & .05  & .07   \\
Assistance      & .14** & .07  & .06   & -.09* & .01  & .09*  \\
Chance          & .09*  & .09* & .05   & -.06 & .01  & .21** \\
Immersion       & .04   & -.03 & .07   & -.03 & -.00 & .01   \\
\hline
\end{tabular}
\caption{Pearson correlations between HEXAD player types and gamification element preferences. 
Values represent correlation coefficients ($r$). 
* $p < .05$, ** $p < .01$ (FDR-corrected within player-type family).}
\label{tab:hexad_correlations}
\end{table}

\subsection{Personality Preferences}
Correlation analyses between Big Five traits and gamification preferences revealed predominantly small effect sizes. No trait showed consistently strong associations across elements.

\textit{Agreeableness} was most clearly associated with \textit{Cooperation}, reflecting its prosocial orientation, with smaller associations for \textit{Altruism}, \textit{Progression}, and \textit{Feedback}.  

\textit{Extraversion} correlated positively with socially and reward-oriented elements, including \textit{Virtual Economy}, \textit{Cooperation}, \textit{Assistance}, and \textit{Chance}.  

\textit{Conscientiousness} showed its strongest associations with \textit{Time Pressure}, \textit{Feedback}, and \textit{Assistance}, aligning with structure- and performance-related preferences.  

\textit{Neuroticism} demonstrated only limited associations, including a positive relationship with \textit{Customization} and a negative association with \textit{Competition}, both small in magnitude.  

\textit{Openness} showed comparatively broader associations, particularly with \textit{Customization} and \textit{Immersion}, consistent with its creativity- and exploration-related characteristics.

Although most correlations were small, several theoretically coherent relationships emerged, such as \textit{Agreeableness–Cooperation}, \textit{Extraversion–Chance}, \textit{Conscientiousness–Assistance}, and \textit{Openness–Customization}. Overall, personality traits explained limited but conceptually meaningful variation in gamification preferences.

\begin{table}[ht]
\centering
\small
\begin{tabular}{l c c c c c}
\hline
\textbf{Element} & \textbf{Agree.} & \textbf{Extra.} & \textbf{Consc.} & \textbf{Neuro.} & \textbf{Open.} \\
\hline
Altruism        & .10*  & .07   & .00   & -.02  & .13** \\
Challenge       & .02   & -.00  & .05   & -.07  & .06   \\
Incentive       & .03   & -.03  & .11*  & -.06  & .04   \\
Competition     & -.04  & .05   & -.01  & -.09* & -.12* \\
Customization   & .03   & .03   & .04   & .13** & .24** \\
Time Pressure   & .02   & .03   & .15** & -.05  & -.03  \\
Progression     & .10*  & -.03  & .06   & .03   & .03   \\
Feedback        & .10*  & -.04  & .14** & -.02  & .06   \\
Virtual Economy & -.01  & .18** & .05   & -.00  & .08   \\
Cooperation     & .20** & .15** & .02   & .01   & -.01  \\
Assistance      & -.00  & .11*  & .18** & .00   & .07   \\
Chance          & -.02  & .20** & .08   & -.01  & .03   \\
Immersion       & .08   & .02   & .02   & .04   & .16** \\
\hline
\end{tabular}
\caption{Pearson correlations between Big Five traits and gamification element preferences. 
Values represent correlation coefficients ($r$). 
* $p < .05$, ** $p < .01$ (FDR-corrected within personality test family).}
\label{tab:bigfive_correlations}
\end{table}

\subsection{Age Preferences}
\begin{figure} [H]
  \centering
  \includegraphics[width=\textwidth]{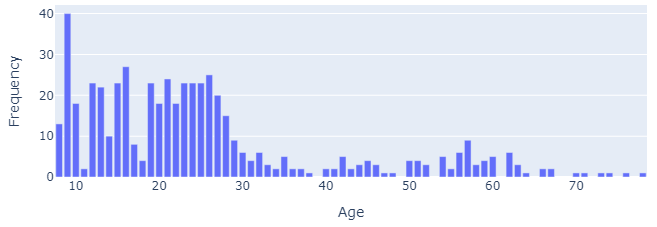}
  \caption{Age Distribution}
  \label{fig:AgeDistributionAll}
\end{figure}

Figure~\ref{fig:AgeDistributionAll} illustrates the age distribution of participants. The sample is skewed toward younger individuals, reflecting the target population of digital learning environments. Distinct peaks among minors result from whole-class participation (grades 3, 4, 7, and 10), while a second concentration appears in the university-age range due to recruitment among students.

Table~\ref{tab:age_correlations} summarizes associations between Age and gamification preferences. Given the non-normal age distribution, Spearman’s rank correlation was used as the primary measure; Pearson correlations were computed as robustness checks and yielded comparable patterns. Negative coefficients indicate stronger preferences among younger participants, whereas positive values indicate stronger preferences among older participants.

No meaningful associations were observed for \textit{Competition}, \textit{Progression}, \textit{Assistance}, or \textit{Immersion}. Several elements, however, showed consistent age-related trends. Preferences for \textit{Incentive}, \textit{Customization}, \textit{Virtual Economy}, \textit{Chance}, and \textit{Altruism} decreased with increasing age. In contrast, \textit{Time Pressure}, \textit{Feedback}, and \textit{Challenge} became more positively rated among older participants. 

Although effect sizes were small to moderate, these trends were statistically stable after FDR correction and conceptually coherent. Younger participants tended to prefer playful and reward-oriented elements, whereas older participants expressed relatively stronger preferences for structured and performance-related design features.

Overall, Age emerged as one of the most consistent predictors of gamification preferences in this study.

\begin{table}[ht]
\centering
\small
\begin{tabular}{l c}
\hline
\textbf{Gamification Element} & \textbf{Spearman $r_s$} \\
\hline
Altruism        & -.13** \\
Challenge       &  .13** \\
Incentive       & -.15** \\
Competition     & -.04   \\
Customization   & -.18** \\
Time Pressure   &  .15** \\
Progression     &  .02   \\
Feedback        &  .16** \\
Virtual Economy & -.22** \\
Cooperation     & -.11*  \\
Assistance      &  .02   \\
Chance          & -.18** \\
Immersion       & -.04   \\
\hline
\end{tabular}
\caption{Spearman correlations between Age and gamification preferences. 
Negative values indicate stronger preferences among younger participants; positive values indicate stronger preferences among older participants. 
* $p < .05$, ** $p < .01$ (FDR-corrected).}
\label{tab:age_correlations}
\end{table}

\subsection{Age Groups}

Since the correlation analysis suggested that age-related preferences may not follow a purely linear trajectory, an additional clustering analysis was performed. 
Using k-means with the elbow method, six clusters were identified as optimal. 
To preserve age proximity within clusters, age values were weighted accordingly. K-means clustering was applied using age and preference ratings across all gamification elements. To emphasize proximity in chronological age, an additional weighted age feature (age × 5) was introduced before clustering. The elbow method indicated six clusters as optimal.
This resulted in the following groups: \textit{[8–10]}, \textit{[11–17]}, \textit{[18–28]}, \textit{[29–38]}, \textit{[39–48]}, and \textit{[49–80]}.

ANOVA results confirmed significant differences across these clusters ($p < .05$, $\eta^2 \approx .045$). 
Post-hoc tests revealed no meaningful differences among the adult groups \textit{[29–38]}, \textit{[39–48]}, and \textit{[49–80]}, indicating stable preferences within mature learners.
Both Pearson and Spearman correlations were tested, with results noted where significance was confirmed by only one method.
The analysis confirmed largely linear trends: younger groups (\textit{[8–10]} and \textit{[11–17]}) rated \textit{Altruism}, \textit{Incentive}, \textit{Customization}, and \textit{Virtual Economy} higher, while older groups favored \textit{Time Pressure} and \textit{Feedback}. 
\textit{Challenge} showed an upward trend, particularly pronounced in the \textit{[18–28]} group, whereas \textit{Chance} declined with age. Both effects were confirmed only by Spearman’s test.
However, three elements displayed non-linear trajectories. \textit{Feedback} was unexpectedly high among the youngest group but lowest among \textit{[11–17]}, producing a U-shaped curve.
Similarly, \textit{Assistance} dipped in mid-adolescence and early adulthood but increased again among older participants, with significance found only in Pearson’s test.
\textit{Cooperation} also followed a U-shaped pattern, being stronger among the youngest and oldest participants, with significance supported only by Spearman.
Additionally, no meaningful age effects were found for \textit{Competition}, \textit{Immersion}, and \textit{Progression}, consistent with the correlation analysis.
In sum, clustering confirms that while several elements follow clear age-related linear trends, others—most notably \textit{Feedback}, \textit{Assistance}, and \textit{Cooperation}, exhibit distinct non-linear patterns that simple correlations would obscure.

\subsection{Learning Activity Task}

This section analyzes preferences across the six \textit{Learning Activity Tasks} (LATs) derived from Bloom’s revised taxonomy: \textit{Remember}, \textit{Understand}, \textit{Apply}, \textit{Analyze}, \textit{Evaluate}, and \textit{Create}. 
As all participants evaluated each LAT, a repeated-measures design was applied. Sample sizes were high and balanced across tasks.

Repeated-measures ANOVAs revealed statistically significant differences across LATs for all gamification elements after FDR correction, although effect sizes ($\eta^2$) were small. \textit{Feedback} showed the weakest variation across task types.

Post-hoc comparisons (Tukey-adjusted) indicated systematic patterns. Lower-order tasks such as \textit{Remember} and \textit{Apply} were more strongly associated with reward-based and performance-oriented elements, including \textit{Incentive}, \textit{Challenge}, \textit{Competition}, \textit{Progression}, and \textit{Time Pressure}. 

Tasks requiring deeper comprehension and reflection, such as \textit{Understand}, \textit{Analyze}, and \textit{Evaluate}, were comparatively more aligned with cooperative and supportive elements, including \textit{Cooperation}, \textit{Assistance}, and \textit{Immersion}.

Notably, \textit{Remember} and \textit{Apply} did not differ significantly from one another, nor did \textit{Analyze} and \textit{Understand}, suggesting shared suitability for similar gamification strategies.

In contrast, \textit{Create} and \textit{Evaluate} showed comparatively fewer strong positive associations overall, indicating that gamification effects may be more context-sensitive in tasks emphasizing reflection, synthesis, or creativity.

Overall, results suggest that the suitability of gamification elements depends substantially on the cognitive demands of the learning task.

\begin{table}[!htb]
\centering
\begin{tabular}{|l|c|c|c|c|c|c|}
\hline
~ & \textit{Remember} & \textit{Understand} & \textit{Apply} & \textit{Analyze} & \textit{Evaluate} & \textit{Create} \\ \hline
\textit{Altruism}      & - &   & - & - & + & - \\
\textit{Assistance}    & - & + & - & + &   &   \\
\textit{Challenge}     & + &   & + &   & - &   \\
\textit{Competition}   & + &   & + &   & - &   \\
\textit{Cooperation}   & - & + &   & + & + &   \\
\textit{Feedback}      & - & + &   & + &   &   \\
\textit{Immersion}     & - & + & - &   & - & + \\
\textit{Incentive}     & + &   & + &   &   &   \\
\textit{Progression}   & + &   & + &   & - & - \\
\textit{Time Pressure} & + &   & + &   &   &   \\ \hline
\end{tabular}
\caption{Significant differences in element preferences across Learning Activity Tasks (repeated-measures ANOVA with Tukey-adjusted post-hoc comparisons). 
“+” indicates significantly higher ratings relative to other tasks; “–” indicates significantly lower ratings. Only statistically significant contrasts are shown.}
\centering
\small + = positive effect, - = negative effect
\label{tab:RecommendationLAT}
\end{table}

\subsection{Evaluating Feature Importance Across Parameters}

An exploratory \textit{Random Forest regression model} was applied to estimate relative feature importance. The model was used for interpretative ranking of predictors rather than predictive optimization. \textit{Learning Activity Tasks} were excluded due to their within-subject evaluation structure. Only cases with complete data across all predictors were included to ensure consistent input features.

\begin{figure} [H]
  \centering
  \includegraphics[width=\textwidth]{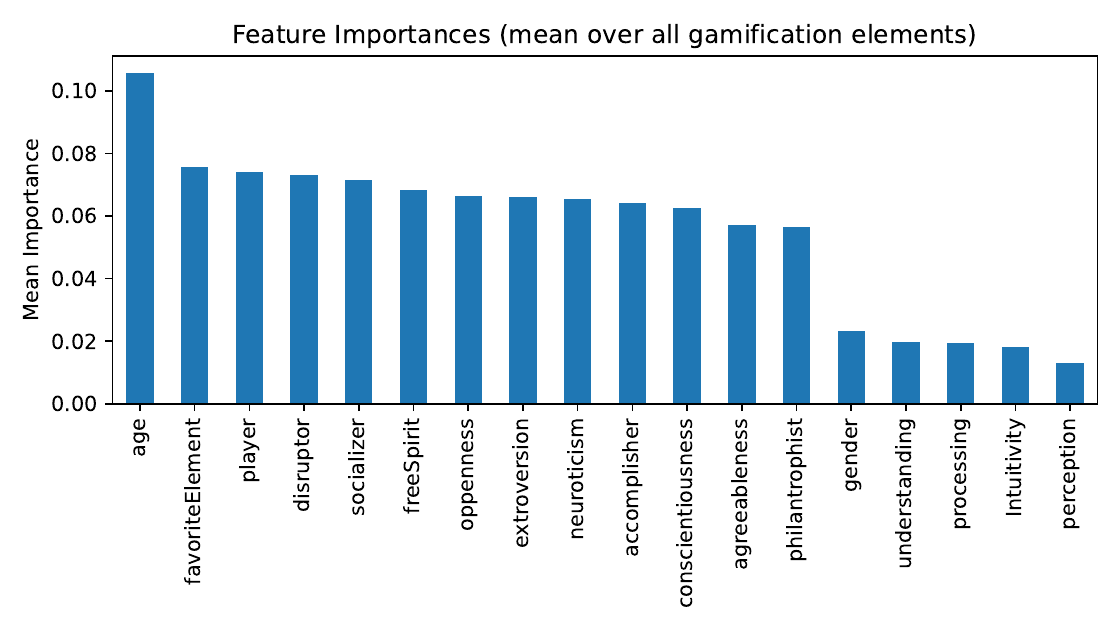}
  \caption{Feature Importance for Predicting Gamification Element Preferences}
   \label{fig:RandomForestResult}
\end{figure}

The model was used to estimate feature importance scores for predicting gamification element preferences. As shown in Figure~\ref{fig:RandomForestResult}, \textit{Age} emerged as the most influential predictor within this modeling framework. \textit{Player Type} and \textit{Personality Traits} showed moderate and comparable importance, whereas \textit{Gender} and \textit{Learning Style} contributed only marginally to predictive performance.

These results are consistent with the preceding inferential analyses and should be interpreted as exploratory rather than causal. Overall, the findings suggest that Age represents the most informative tailoring parameter in this dataset, followed by motivational and personality-related factors.



%% file: Chapters/6.DiscussionConclusion.tex
\section{Discussion}
This study provides a large-scale empirical examination of how individual user traits and learning activity contexts relate to preferences for gamification elements. By jointly analyzing age, gender, HEXAD player types, Big Five personality traits, learning styles, and Bloom-based learning tasks within a unified taxonomy, the results offer a differentiated perspective on motivational diversity in digital learning environments.

Across all parameters, \textit{Age} emerged as the most consistent predictor of gamification preferences. Correlational and cluster-based analyses showed that younger participants tended to favor reward-oriented and exploratory elements such as \textit{Incentive}, \textit{Customization}, and \textit{Virtual Economy}, whereas older participants preferred structured and performance-related elements including \textit{Challenge}, \textit{Time Pressure}, and \textit{Feedback}. These patterns were stable across Pearson and Spearman analyses and partially confirmed in age-cluster comparisons. While effect sizes were generally small to moderate, their consistency across multiple statistical approaches suggests that age represents a meaningful design dimension in gamified learning systems.

Gender differences were statistically significant for several elements, but effect sizes remained small. Women showed stronger preferences for supportive and creative elements such as \textit{Assistance} and \textit{Feedback}, whereas men expressed higher preference for \textit{Competition}. However, given the modest magnitude of these effects, gender alone appears insufficient as a primary personalization criterion. These findings reinforce previous work highlighting motivational heterogeneity while cautioning against overreliance on demographic simplifications.

The HEXAD player type framework demonstrated clearer and theoretically coherent patterns. \textit{Socializers} showed the strongest association with \textit{Cooperation}, \textit{Accomplishers} with \textit{Challenge}, and \textit{Players} with \textit{Incentive}, aligning closely with theoretical expectations of the model. Although most correlations were weak in magnitude, several associations were stable across parametric and non-parametric tests. This suggests that archetype-based profiling may offer more behaviorally relevant personalization signals than isolated demographic traits.

Big Five personality traits yielded weaker but interpretable patterns. \textit{Agreeableness} correlated with \textit{Cooperation}, \textit{Conscientiousness} with \textit{Time Pressure} and \textit{Assistance}, and \textit{Openness} with \textit{Customization} and \textit{Immersion}. Notably, no meaningful associations emerged for \textit{Challenge}, indicating that some elements may function independently of broad personality dimensions. Overall, personality appears to contribute incremental explanatory value but does not dominate preference formation.

Beyond individual traits, learning activity context significantly influenced perceived suitability of gamification elements. Lower-order tasks (e.g., \textit{Remember}, \textit{Apply}) aligned more strongly with competitive and reward-based elements such as \textit{Competition}, \textit{Progression}, and \textit{Incentive}, whereas higher-order or reflective tasks (e.g., \textit{Evaluate}, \textit{Understand}) were more compatible with \textit{Cooperation} and \textit{Assistance}. Although effect sizes were small, the systematic differentiation across Bloom’s levels supports the notion that gamification should be task-sensitive rather than uniformly applied.

Interestingly, certain elements such as \textit{Challenge}, \textit{Progression}, and \textit{Feedback} showed relatively broad acceptance across traits and contexts. These elements may serve as “stable anchors” within adaptive systems, providing baseline motivational structure while more dynamic elements are adjusted based on user and task characteristics.

\subsection*{Implications for Design}

The findings suggest that effective gamified learning environments should integrate both trait-based and task-based adaptation. Age appears to be the strongest high-level segmentation variable, particularly when designing for mixed-age audiences. Player type profiling may provide additional refinement, especially when selecting between competitive, cooperative, or reward-focused mechanics. Learning task classification further enables dynamic adjustment of gamification layers depending on cognitive demand.

Rather than embedding fixed game mechanics, systems may benefit from modular gamification architectures in which elements can be selectively activated or emphasized. Such modularity supports recommender-based approaches and AI-driven orchestration mechanisms that adapt motivational design to both learner characteristics and instructional context.

\section{Limitations}
Several limitations should be considered when interpreting the results. First, the study relies on self-reported preferences rather than observed behavioral engagement or learning outcomes. Expressed preferences do not necessarily translate into sustained motivation, effective learning behavior, or measurable performance gains. Future studies should therefore complement survey data with experimental or longitudinal designs.

Second, although efforts were made to simplify descriptions for younger participants, certain abstract gamification concepts—such as \textit{Immersion}, \textit{Customization}, and \textit{Time Pressure}—may not have been fully understood by all respondents. In addition, some participants may not have recognized that the task-based questions referred to distinct learning activities, potentially introducing response noise.

Third, the sample, while large, is not fully representative. Recruitment through schools, universities, and social media resulted in an overrepresentation of academically oriented and digitally active participants. Older adults and individuals with limited exposure to digital learning environments may therefore be underrepresented. Moreover, subgroup sizes were imbalanced in some parameters, particularly for \textit{Disruptors} in the HEXAD model and \textit{Verbal} learners in the learning-style dimension, limiting statistical power for those groups.

Fourth, the online survey format introduces uncontrollable factors such as distractions during completion and differences in familiarity with gamified systems. While data cleaning procedures were applied, complete control over multiple submissions or inattentive responding cannot be guaranteed.

Fifth, Likert-scale ratings were treated as interval data for parametric analyses. Although this is common practice in large-sample survey research, it represents a methodological approximation. Complementary non-parametric tests were therefore conducted to enhance robustness.

Finally, the taxonomy of gamification elements reflects conceptual and operational choices. While systematically derived, alternative categorizations may yield different relational patterns. Replication across diverse cultural contexts and with alternative operationalizations is needed to further validate the findings.

\section{Conclusion}
This study provides a comprehensive empirical analysis of how user characteristics and learning activity contexts relate to preferences for gamification elements in digital education. Across multiple parameters, including age, player type, personality traits, gender, learning styles, and Bloom-based learning tasks, systematic but mostly small-to-moderate effects were observed.

Age emerged as the most consistent predictor, followed by player type and personality traits, whereas gender and learning styles showed comparatively weaker associations. Importantly, task context significantly influenced the perceived suitability of gamification elements, reinforcing that motivational design should be aligned with pedagogical intent rather than applied uniformly.

The findings suggest that gamification effectiveness cannot be attributed to universally motivating mechanics. While elements such as \textit{Challenge}, \textit{Progression}, and \textit{Feedback} showed relatively broad acceptance, many preferences varied meaningfully across age groups, motivational archetypes, and task types. These results support the development of adaptive and modular gamification architectures that allow selective activation of elements depending on learner characteristics and instructional context.

Future research should extend this work by examining behavioral outcomes, longitudinal effects, and interactions among multiple traits. Additionally, refining conceptual distinctions within gamification elements (e.g., different forms of time-related constraints) may yield deeper insights into motivational mechanisms.

Overall, this study contributes to a more differentiated and evidence-informed understanding of gamification design. Rather than advocating one-size-fits-all solutions, the results highlight the importance of personalization grounded in empirical analysis of user diversity and learning context.